# Creating an Isotopically Similar Earth-Moon System with Correct Angular Momentum from a Giant Impact

Bryant M. Wyatt[1*], Jonathan M. Petz[1,2], William J. Sumpter[1,2], Ty R. Turner[1,2], Edward L. Smith[1,2], Baylor G. Fain[3], Taylor J. Hutyra[1,3], Scott A. Cook[1], Michael F. Hibbs[3], Shaukat N. Goderya[3,4]

[1]Department of Mathematics, Tarleton State University, Stephenville, TX 76402.

[2]Department of Engineering and Computer Science, Tarleton State University, Stephenville, TX 76402.

[3]Department of Chemistry, Geoscience, and Physics, Tarleton State University, Stephenville, TX 76402.

[4]Program for Astronomy Education & Research, Tarleton State University, Stephenville, TX 76402.

## Abstract

The giant impact hypothesis is the dominant theory explaining the formation of our Moon. However, its inability to produce an isotopically similar Earth-Moon system with correct angular momentum has cast a shadow on its validity. Computer-generated impacts have been successful in producing virtual systems that possess many of the physical properties we observe. Yet, addressing the isotopic similarities between the Earth and Moon coupled with correct angular momentum has proven to be challenging. Equilibration and evection resonance have been put forth as a means of reconciling the models. However, both were rejected in a meeting at The Royal Society in London. The main concern was that models were multi-staged and too complex. Here, we present initial impact conditions that produce an Earth-Moon system whose angular momentum and isotopic properties are correct. The model is straightforward and the results are a natural consequence of the impact.

## Introduction and Background

The canonical giant impact hypothesis (*1-3*) states that an impactor, roughly the size of Mars, obliquely struck proto-Earth. The dense iron cores of proto-Earth and the impactor coalesced to form Earth's core. A substantial amount of the outer silicate material was ejected into orbit around the newly formed Earth and coalesced into our Moon. Computer-generated impacts were successful in producing the Earth with a circumplanetary disk of debris that was low in iron content and had enough mass to produce the Moon (*4-7*). The angular momenta of the computer-generated Earth-Moon systems were also close to the measure value. Separate simulations, starting with a circumplanetary disk of debris, showed how orbiting material could coalesce into the Moon (*8, 9*).

In these simulations, a large proportion of the debris within the circumplanetary disk came from the relatively small impactor (*10, 11*). However, detailed analysis of the Moon samples reveal striking isotopic similarities between the Earth and Moon (*10-13*), making it unlikely that the Moon could be composed predominantly of impactor material. Attempts have been made to reconcile this issue, but all have led to complex multi-stage models.

Stevens and Pahlevan (*14*) proposed adding equilibration to the existing models to resolve the isotopic concerns. Turbulent mixing between the Earth and the circumplanetary disk could have created a Moon that was isotopically similar to Earth. Other researchers modified the canonical impact to create models that would, from the impact, produce an Earth and a circumplanetary disk that were isotopically similar. This could be done in two ways: – (i) create a disk that was composed predominantly of proto-Earth material, or (ii) produce an Earth and a disk that were both composed of relatively equal amounts of material from both impactor and target. Cuk and Stewart (*15*) used a small impactor colliding with a rapidly spinning proto-Earth to create a disk composed predominantly of material from proto-Earth. Canup (*16*) used impactors of equal size in an off-centered collision to create an Earth and disk composed of equal amounts of material from both impactors. Eiland et al. (*17*) used impactors of equal size in an off-centered collision, coupled with impactor spins, to create an Earth and a Moon composed of equal parts from both impactors. Reufer et al. (*18*) used a hit-and-run scenario that produced a disk composed predominantly of material from proto-Earth. These variations were successful in addressing the isotopic concerns, but all produced an Earth-Moon system with excess angular momentum. Cuk and Stewart (*15*) proposed using evection resonance between the Moon and the Sun as a mechanism to remove the excess angular momentum.

In the summer of 2013, The Royal Society called a meeting solely to discuss the formation of the Moon (*19*). In this meeting, evection resonance and equilibration were both rejected as viable means of removing the deficiencies from giant impact models (*20, 21*). The following is an excerpt written by Robin Canup (*22*), associate vice-president of the Planetary Directorate of the Southwest Research Institute in Boulder Colorado, after the meeting:

> The main concern being that the solutions had become too complex and unnatural. It remains troubling that all of the current impact models invoke a process after the impact to effectively erase a primary outcome of the event — either by changing the disk's composition through mixing for the canonical impact, or by changing Earth's spin rate for the high-angular-momentum narratives.

> Sequences of events do occur in nature, and yet we strive to avoid such complexity in our models. We seek the simplest possible solution, as a matter of scientific aesthetics and because simple solutions are often more probable. As the number of steps increases, the likelihood of a particular sequence decreases. Current impact models are more complex and seem less probable than the original giant-impact concept.

### The Model and Collision Scenarios

Here, we present a model that produces an isotopically similar Earth-Moon system with the correct angular momentum. This is done in a single, straightforward simulation. Isotopic similarity and correct angular momentum result naturally from the collision. Neither evection resonance nor equilibration are needed.

Observing the obstacles encountered in prior work, focus was placed on producing a model where results evolved solely from the impact. This dictated an extensive search of the initial parameter space be performed. To accomplish this, a model was sought that was computationally inexpensive and easily parallelized to run on modern graphics processing units. Eiland's model (17) was ideal for this purpose because it adopts a simplified approach to thermodynamics, yet is otherwise physically realistic. Another advantage of Eiland's model is that it frequently produces a fully formed Moon, whereas other models typically only produce a circumplanetary disk of debris.

Consider the scenarios depicted in Fig. 1. There are three contributors to the angular momentum of the system — the spin of impactor 1, the spin of impactor 2 and, the rotation of the entire system about its center of mass created by the off-center trajectories. We denote them $L_1$, $L_2$, and $L_{cm}$ respectively. An impactor is said to spin into the collision if its spin is opposed to $L_{cm}$, and out of the collision if its spin aligns with $L_{cm}$. Kinetic energy comes from impactor spins and the closing velocity of the impact. The key is to produce sufficiently high kinetic energy while controlling angular momentum.

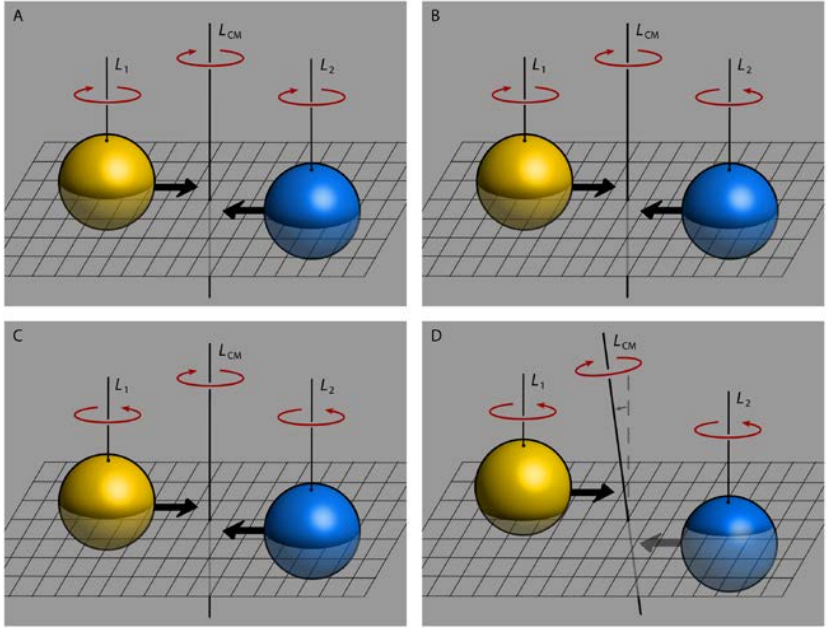

**Fig. 1. Collision scenarios.** The black arrows represent the impactors' initial velocities. The red arrows represent the impactors' initial spins. Impactor 2 (blue) has an initial position towards the reader. In scenario D, the impactors' equatorial planes are also orthogonally separated.

Fig. 1A shows a coplanar, off-centered collision, where both impactors spin out of the collision. This produced smooth and predictable results. By adjusting the spin rates, initial linear velocities, and the off-center displacement, the composition and size of the resulting Moon could be controlled. However, all runs that released enough material into orbit to create the Moon suffered from excess angular momentum (typically a factor of at least two greater than the true Earth-Moon system). Here $L_1$, $L_2$, and $L_{cm}$ augment each other, giving little latitude to reduce angular momentum without reducing kinetic energy.

Fig. 1B shows a coplanar, off-centered collision where impactor 2 spins into the collision and impactor 1 spins out of the collision. This allowed the reduction of angular momentum while keeping the kinetic energy of the system high, because $L_2$ was opposed to $L_1$ and $L_{cm}$. However, the released material was almost exclusively from the impactor spinning out of the collision. Hence, this type impact did not satisfy the isotopic similarity condition.

Fig. 1C shows a coplanar, off-centered collision where both impactors spin into the collision. This allowed easy reduction of angular momentum while keeping kinetic energy high, because both $L_1$ and $L_2$ were opposed to $L_{cm}$. The inward rotations limited lateral escape of material. If the kinetic energy was sufficiently high, the collision was violent and chaotic, and large amounts of material were ejected above and below the collision plane. Because the impact was coplanar and symmetric, the majority of the ejected material was released orthogonal to the collision plane, and returned back to Earth. This resulted in an insufficient amount of orbital debris. However, a positive result was that the debris was composed of equal amounts from both impactors.

Fig. 1D shows a collision identical to Fig. 1C, except that the equatorial planes of the impactors are orthogonally offset. This produced an additional contribution to angular momentum not seen in Fig. 1A-C, which tilted the axis along which material was primarily ejected. This gave the ejected material an angular component which increased the amount of orbital debris. The orbital debris was massive enough to create the Moon, iron-poor, and composed of relatively equal amounts of material from both impactors. The angular momentum of the resulting system could also be controlled while keeping kinetic energy high. This type impact satisfied all necessary conditions.

## Simulation

If kinetic energy is high, and total angular momentum is set close to the observed value of the Earth-Moon system, the type collision depicted in Fig. 1D consistently produces favorable results. To demonstrate this, a simulation is presented where the initial impactor spins and velocities were selected from Cuk and Stewart's 2012 paper (15). The remaining initial conditions were then set so the angular momentum for the entire system would be close to what is measured for the Earth-Moon system today. Cuk and Stewart present a series of runs where proto-Earth is spinning with a 2.3 hour period and is hit with impactors whose ratio of impact velocity to mutual escape velocity ($v_{imp}/v_{esc}$) range from 1 to 3. In the simulation presented here, both impactors were given a 2.3 hour period and an impact ratio $v_{imp}/v_{esc}$ equal to 2. See Table 1 for a full list of initial conditions.

**Table 1. Initial Values for Impactors.**

| Parameter | Impactor 1 (yellow) | Impactor 2 (blue) |
|---|---|---|
| Center of mass | (-44500, -200, -1810) km | (44500, 200, 1810) km |
| Linear velocity | (8.8, 0.0, 0.0) km/s | (-8.8, 0.0, 0.0) km/s |
| Angular velocity | (0.0, 0.43, 0.0) rev/h | (0.0, 0.43, 0.0) rev/h |
| % of Earth's mass | 55 | 55 |
| % Iron material by mass | 26 | 26 |
| % Silicate material by mass | 74 | 74 |
| # of computational elements | 65,536 | 65,536 |

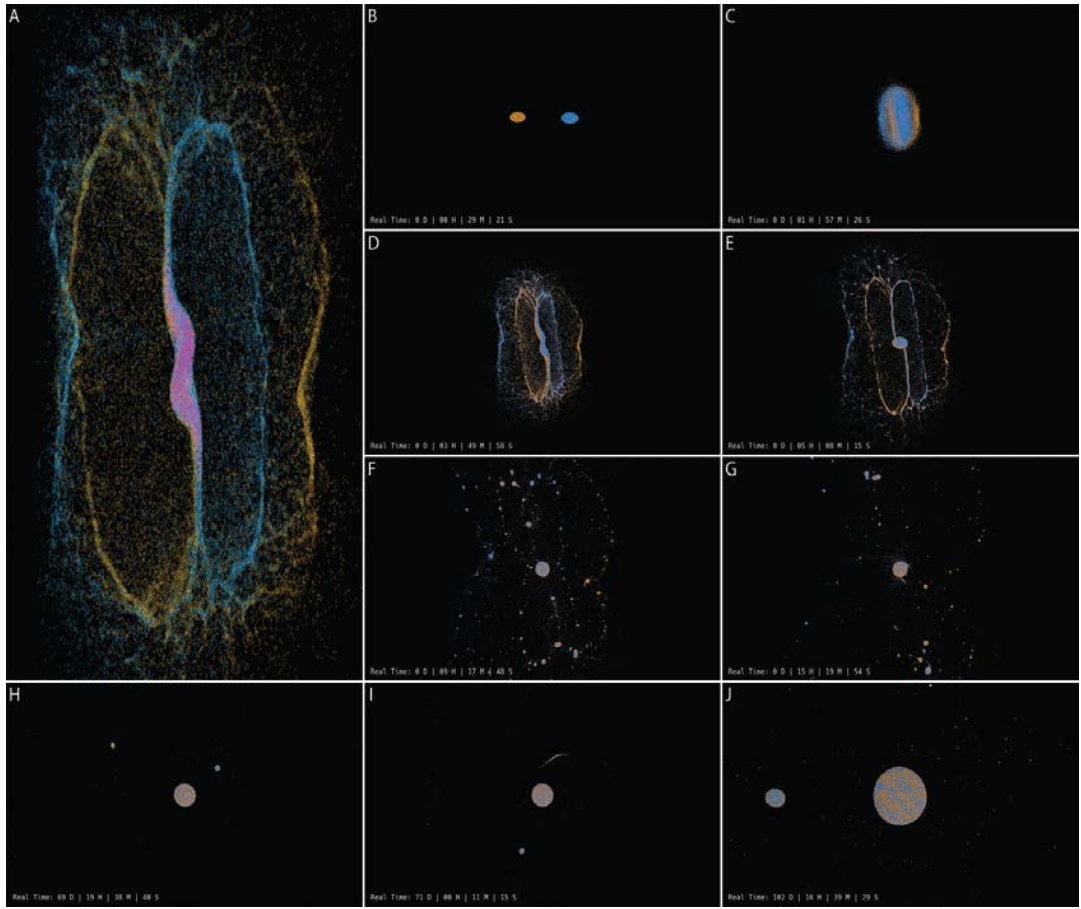

**Fig. 2. Inwardly rotating, off-planar, off-center collision.** Impactor 1 (left, silicate-yellow, iron-red); impactor 2 (right, silicate-blue, iron-purple). The impactors have parallel equatorial planes that are orthogonal to the field of view. The planes are slightly offset, with plane 1 below plane 2. This side view shows the orthogonal release of material. (A) The apex of the collision with the iron elements enhanced. (B-G) Snap shots 0.49, 1.95, 3.82, 5.13, 9.28, and 15.32 hours into the collision. (H) Two dominant moons in orbit. (I) Destruction of the companion moon at 1704.18 hours. (J) Earth and Moon at 2464.48 hours at which time the statistics of the Earth-Moon system were gathered. Videos of the simulation are presented in the supplementary material (movie S1-S3).

In this simulation, the material released to form the orbital debris was iron free, composed of 59.3% material from impactor 1 and 40.7% from impactor 2. The resultant Earth was composed of 49.5% material from impactor 1 and 50.5% material from impactor 2. The angular momentum of the resultant Earth-Moon system was $3.5677 \times 10^{28}$ (kg·km$^2$/s) which is only 2.7% off today's measured value of $3.4738 \times 10^{28}$ (kg·km$^2$/s). The ratio of the size of the Earth to that of the Moon was 84.65 which is only 4.1% off the actual ratio of 81.28. Additional results are presented in Table 2. The collision is illustrated in Fig. 2.

**Table 2. Simulation Results of the Earth and Moon 2464.48 hours into the simulation.**

| Parameter | Earth | Moon |
| --- | --- | --- |
| Mass | $6.3106 \times 10^{24}$ kg | $7.4551 \times 10^{22}$ kg |
| Iron material elements from impactor 1 | 8,479 | 0 |
| Iron material elements from impactor 2 | 8,478 | 0 |
| Silicate material elements from impactor 1 | 54,582 | 497 |
| Silicate material elements from impactor 2 | 53,457 | 1253 |
| Mixing ratio (body 2 elements)/(body 1 elements) | 0.982 | 2.521 |
| %Iron material by mass | 27.1 | 0.0 |
| %Silicate material by mass | 72.9 | 100.0 |
| Axial tilt off the average collision plane | 15.83 degrees | |

## Conclusion and Observations

In summary, it was observed that if two roughly half-Earth-sized impactors, rotating in close to the same plane, collide in a high-energy, off-centered impact — where both impactors spin into the collision — an iron-deficient Moon composed of large percentages of material from both impactors can form. In addition, the resulting Earth-Moon system can have an angular momentum comparable to what we see today. The collisions are violent, but this is in line with recent papers on high-energy impact and vigorous mixing during Moon-forming events (*23, 24*).

One interesting result of this type of collision is that the equatorial plane of the Earth, and the average orbital plane of the ejected material, are nearly orthogonal. This was initially surprising. However, the Moon's orbit today is substantially off the equatorial plane (*25, 26*). Tidal forces produced by Earth's rapid rotation are transferring energy to the Moon, causing it to move outward, parallel to the Earth's equatorial plane (*27-29*). This process is constantly reducing the angle between the Moon's orbit and the equatorial plane. Hence, the Moon's off-equatorial orbit is produced naturally in the model and will evolve toward what we see today.

Another interesting result, is that two dominant moons usually emerge and orbit the Earth for an extended period of time. On most occasions, one of these moons drives the other moon inside Earth's Roche limit where it is ripped apart by Earth's gravity. However, in rare cases, the two moons collide, which is in line with companion moon theories of the differences between the near and far side of the Moon (*30*).

## Acknowledgments

We thank: NVIDIA and Mellanox Technologies for donation of hardware; Tarleton State University's high performance computing lab for space; Tarleton State University's Office of Faculty Research, Office of Student Research, and The Program for Astronomy Education and Research for funding; John H. Gresham and Robert W. Muth for insightful remarks and suggestions; and special thanks to The Wyatt Foundation for supporting student travel.